\documentclass{emulateapj}

\newcommand{\gr}{$\gamma$-ray \,}
\newcommand{\grs}{$\gamma$-rays \,}

\shorttitle{Nature of the \gr emission of Tycho's SNR}
\shortauthors{Berezhko et al.}

\begin{document}

\title{The nature of gamma-ray emission of Tycho's supernova remnant}
 
\author{E.G. Berezhko\altaffilmark{1},
        L.T. Ksenofontov\altaffilmark{1},
    and H.J. V\"olk\altaffilmark{2}
}

\altaffiltext{1}{Yu.G. Shafer Institute of Cosmophysical Research and Aeronomy,
                     31 Lenin Ave., 677980 Yakutsk, Russia}
\altaffiltext{2}{Max Planck Institut f\"ur Kernphysik,
                Postfach 103980, D-69029 Heidelberg, Germany}

\email{berezhko@ikfia.ysn.ru}  


\begin{abstract}
The nature of the recently detected HE and VHE \gr emission of
Tycho's supernova remnant (SNR) is studied.
A nonlinear kinetic theory of cosmic ray (CR) acceleration in supernova
remnants (SNRs) is employed to investigate the properties of Tycho's SNR and
their correspondence to the existing experimental data, taking into account
that the ambient interstellar medium (ISM) is expected to be clumpy.
It is demonstrated that the overall steep \gr spectrum observed can be
interpreted as the superposition of two spectra produced by the CR proton
component in two different ISM phases: The first \gr component, extending up
to about $10^{14}$~eV, originates in the diluted warm ISM, whereas the second
component, extending up to 100~GeV, comes from numerous dense, small-scale
clouds embedded in this warm ISM.
Given the consistency between acceleration theory and the observed properties
of the nonthermal emission of Tycho's SNR, a very efficient production of
nuclear CRs in Tycho's SNR is established. The excess of the GeV \gr emission
due to the clouds' contribution above the level expected in the case of a
purely homogeneous ISM, is inevitably expected in the case of type Ia SNe.
\end{abstract}


\keywords{(ISM:)cosmic rays -- 
                acceleration of particles -- 
                shock waves --
                supernovae individual(Tycho's SNR) -- 
                radiation mechanisms:non-thermal --
                gamma-rays:theory}

\section{Introduction}

The detection of high-energy (HE; $100~\mathrm{MeV} \leq E \leq 100$~GeV) and
very high energy (VHE;$E \geq 100$~GeV) \gr emission from supernova remnants
(SNRs) is extremely important, because it provides direct evidence for the
acceleration of charged particles (atomic nuclei and/or electrons) inside SNRs
to energies that are comparable to those of the gamma rays. Based on such
detections one can hope to eventually confirm the idea that SNRs are indeed the
main source of nuclear cosmic rays (CRs) up to energies of about $10^{17}$~eV
in the Galaxy, as widely expected
\citep{bellucek01,ptuskin03,bv07,vbk08,bkv09}, see however \citet{parizot06}.

However, even when \grs from a SNR are detected, it is not a simple matter to
determine the types of energetic particles that produce them. If the detection
is in TeV emission only, like in the case of SN~1006 up to now
\citep{sn1006hess10}, then one has to find evidence  that the emission has a
substantial $\pi^0$-decay \gr component produced by nuclear CRs, or whether it
is merely inverse Compton (IC) and/or Bremsstrahlung emission produced by
accelerated electrons.

At a Galactic latitude of about 15 degrees, SN~1006 lies about 550~pc above the
Galactic gas disk \citep{winkler03,bocc11}. From its dipolar symmetry in hard
X-rays and TeV gamma rays it appears a good approximation to assume an
interstellar environment that is uniform both in gas density as well as in
magnetic field direction and strength, even though this simplification is still
in need of investigation. The dipolar symmetry can be attributed to a
corresponding characteristic of the injection process of suprathermal nuclei
from the shocked downstream plasma into the acceleration process: such
particles can outrun the shock only if the outer SNR shock is quasi-parallel to
the upstream magnetic field and only then can they participate in the diffusive
shock acceleration process. As argued earlier \citep{vbk03,bkv09} this means
that only approximately 20 percent of the total shock surface will contribute
to nuclear CR production from a type Ia SNR and nonlinear shock-modification
effects like magnetic field amplification in a type Ia SN, whereas about 80
percent of the shock surface correspond to a quasi-perpendicular shock without
significant injection of nuclei and to suppressed acceleration. A
spherically symmetric acceleration model therefore has to be ``renormalized''
in the sense that the integrated particle spectrum obtained in spherical
symmetry needs to be diminished by a factor $\approx 0.2$\footnote{This
  argument should hold also for the probably significantly more complex and
  smaller-scale inhomogeneous but topologically equivalent external magnetic
  field configuration, expected for type Ia SNRs in general and for Tycho's SNR
  in particular \citep{vbk03}.}  Making this correction, the sum of the
multi-wavelength observations and their theoretical interpretation suggests
that nuclear CRs are produced in SN~1006 with high efficiency, as required for
the Galactic CR sources \citep{bkv12}.

With significant detections at GeV as well as at TeV energies of otherwise
simple objects like remnants of type Ia supernovae in the Galactic gas disk,
there is still the question whether the circumstellar environment is uniform in
gas density or not, and what role this non-uniformity plays for the overall
observed \gr spectrum and morphology. This question is discussed here for
Tycho's SNR. The remnant is spatially unresolved in \grs but otherwise very
well studied. The spatially-integrated TeV-spectrum, detected by the VERITAS
array \citep{veritas11}, is quite compatible with a theoretical model for a
type Ia - explosion in a strictly uniform Interstellar Medium (ISM), studied in
detail in a previous paper \citep{vbk08} which is in the following referred to
as VBK08.  However, the recent Fermi Large Area Telescope (Fermi LAT) detection
of high energy \grs above 400~MeV \citep{giordano12} disagrees with this simple
nonlinear model, showing a significant GeV excess. \citet{morlino12} (in the
following referred to as MC12) have attempted to understand this result
assuming a spectrum of CR protons $N\propto \epsilon^{-\gamma}$ with a spectral
index $\gamma = 2.2$. This spectrum is considerably steeper than the spectrum
predicted by VBK08, that implies $\gamma \approx 2$. To obtain this result,
MC12 assumed Bohm diffusion for all the accelerated particles. However, as will
be shown below, such an interpretation contains an internal contradiction: Bohm
diffusion even at the highest particle energies involved is inconsistent with
such a steep proton spectrum.

In this paper a new interpretation of the detected \gr spectrum of Tycho's SNR
will be given. It is based on the expectation that the actual interstellar
medium (ISM) is clumpy instead of being purely homogeneous
\citep{field1965,wolfire2003}.

\section{Homogeneous ISM}

The present form of the solutions of the nonlinear acceleration equations,
considered here \citep[see][for a review]{ber08}, assumes spherical
symmetry. In this approximation it is possible to predict the temporal and
radial evolution of gas density, pressure, and mass velocity, together with
that of the energy spectrum, as well as the spatial distribution of CR nuclei
and electrons, including the properties of their non-thermal radiation.

This theoretical model has been used in detail to investigate Tycho's SNR as
the remnant of a type Ia SN \citep{krause08} in a homogeneous ISM, in order to
compare the results with the existing data (VBK08). It was argued that
consistency of the standard value of stellar ejecta mass
$M_\mathrm{ej}=1.4M_{\odot}$ and a total hydrodynamical explosion energy
$E_\mathrm{sn}=1.2\times 10^{51}$~erg \citep{bad06} with the gas dynamics,
acceleration theory and the existing \gr measurements required the source
distance $d$ to exceed $3.3$~kpc in order to be consistent with the existing
HEGRA upper limit for TeV \gr emission. The corresponding ambient gas number
density $N_\mathrm{g}=\rho/m$ (where $\rho$ is the gas density and $m$ is the
proton
mass) had then to be lower than $0.4$~cm$^{-3}$. On the other hand, the rather
low distance estimates from independent measurements together with internal
consistency arguments of the theoretical model made it even more likely that
the actual \gr flux from Tycho is ``only slightly'' below the HEGRA upper
limit. The strong magnetic field amplification produced by accelerated CRs
implied a mean field strength of $\approx 400~\mu$G \citep[e.g.][]{vbk05}
and as such implied in addition that the \gr flux is
hadronically dominated.  The shock was modified with an overall compression
ratio $\sigma \approx 5.2$ and a subshock compression ratio $\sigma_\mathrm{s}
\approx 3.6$; the latter is consistent with the observed radio index
\citep{rey92}. The differential proton spectral index was $\gamma \approx 2$.

\begin{figure}
\plotone{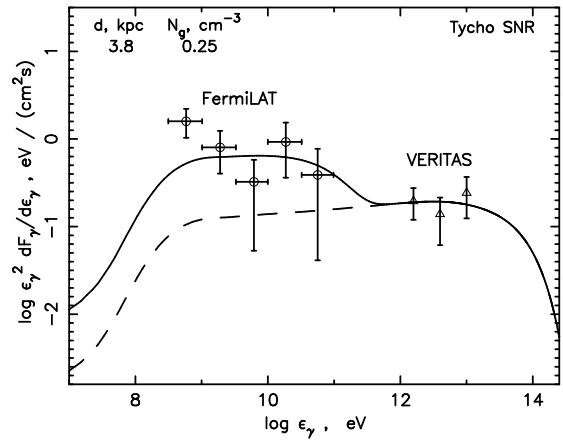}
\caption{Spectral energy distributions of the \gr emission from Tycho's SNR as
functions of \gr energy $\epsilon_{\gamma}$, calculated for a source
distance $d=3.8$~pc, together with the experimental data obtained
by {\it Fermi} and {\it VERITAS}.  Dashed and solid lines represent the
contribution of the warm-phase ISM and the total \gr energy spectrum that
includes the contribution of the clouds (see main text), respectively.}
\label{f1}
\end{figure}

The TeV \gr emission from Tycho detected by VERITAS \citep{veritas11}
corresponds very well to the above expectation. As can be seen from Fig.\ref{f1}
a new \gr spectrum calculated within the kinetic nonlinear theory (shown by the
dashed line) is well consistent with the VERITAS measurements.  This new
calculation was performed following the usual procedure as described in VBK08.
For the proton injection rate $\eta = 3\times10^{-4}$ this is still compatible
with the above-mentioned shock modification and softening of the observed radio
synchrotron emission spectrum. The new distance $d=3.8$~kpc and the
corresponding new ambient ISM number density $N_\mathrm{g}=0.25$~cm$^{-3}$ were
taken in
order to fit the observed TeV \gr emission \citep[see also][]{veritas11}.

However, as mentioned in the Introduction, the \gr spectrum measured by
the Fermi LAT at energies 400~MeV to 100~GeV \citep{giordano12} is considerably
(by a factor 2 to 5) above the value predicted by the kinetic theory (see
Fig.\ref{f1}).  This excess of GeV \gr emission, when compared with the
theoretical predictions, requires a more detailed consideration of this object
and its environment, taking into account new physical factors which had
been hitherto neglected.

\section{Steep CR spectrum?}

The first possibility to explain the complete \gr data might be the search for
a solution with an essentially steeper CR spectrum for all energies above
$\approx 1$~GeV, compared with the hard-spectrum solution discussed in the last
section. Such a soft-spectrum solution was proposed by MC12. These authors also
used a nonlinear approach, assuming a homogeneus circumstellar environment
of the SNR. However, their model differs essentially from the present one in
terms of internal consistency.

For some aspects of the nonlinear problem, that require a consistent
determination of the CR spectrum together with the time-dependent dynamics of
the expanding spherical shock, MC12 substitute a simplified semi-analytical
approach. This is not a principle issue. In addition, there are several basic
differences between the present approach and theirs. First of all, MC12 do not
permit that some part of SN shock should be locally quasi-perpendicular,
where the injection of suprathermal, ionized nuclear gas particles from
downstream is strongly suppressed. Their magnetic field model is
instead topologically equivalent to assuming a monopolar magnetic field.
As a consequence they suggest equally efficient injection/acceleration of CRs
all over the shock surface. Secondly, these authors ignore any nonadiabatic gas
heating within the shock precursor. Therefore the magnetic field pressure in
the resulting model plays a more important role than the gas pressure for the
determination of the shock parameters. On the other hand, numerical simulations
\citep{bell04,zirak08} of the nonresonant instability, caused by CRs within the
precursor, clearly demonstrate significant gas heating that is roughly
consistent with the model considered here \citep[e.g.][]{ber08}.

Under the above assumptions of MC12 it is indeed formally possible to obtain a
solution with a steep CR spectrum, consistent with the existing measurements.
However, apart from the discussed differences between the two approaches, the
scheme of MC12 in addition contains the assumption of Bohm diffusion for all
particles, also for those of the highest energies. For the obtained steep
spectrum this amounts to an internal contradiction which renders such a
solution inconsistent.

To make this clear requires a brief analysis of the general properties of CR
acceleration in spherical blast waves:

The expanding spherical SN shock produces a power law momentum spectrum of
accelerated CRs which extends up to a cutoff momentum determined by the
expression \citep{ber96}
\begin{equation}
\kappa(p_\mathrm{max})=R_\mathrm{s}V_\mathrm{s}/A,
 \label{eq1}
\end{equation}
where $\kappa$ denotes the CR diffusion coefficient, $R_\mathrm{s}$ and
$V_\mathrm{s}$ are the
radius and the speed of the shock respectively, and $A$ is a numerical factor
whose value depends on the shock expansion law. For a rough estimate one can
use the value $A=10$. Note that the above expression is also roughly
  valid for a nonspherical shock, for example for the shock transmitted into
  a dense clump.  In this case half of the mean size of the clump $l$,
i.e. $l/2$, plays the role of the shock radius $R_\mathrm{s}$.  

When CRs are scattered mainly by Alfv\'en waves, as assumed by MC12, the
diffusion coefficient can be written in the form
\begin{equation}
\kappa(p)=\kappa_\mathrm{Bohm}(p)E_\mathrm{B}/E_\mathrm{w}(k=1/\rho_\mathrm{B}),
 \label{eq2}
\end{equation}
where
\begin{equation}
\kappa_\mathrm{Bohm}=\rho_\mathrm{B} v/3
 \label{eq3}
\end{equation}
is the theoretically minimal diffusion coefficient, the so-called Bohm
diffusion coefficient, $v$ and $\rho_\mathrm{B}$ are the particle velocity and
gyroradius, respectively, $E_\mathrm{B} = B^2/(8\pi)$ is the energy density of
the
large-scale magnetic field, and $E_\mathrm{w}(k)$ is the Alfv\'en wave magnetic
energy density per unit logarithm of wave number $k$.

\citet{bell04} found that a non-resonant streaming instability will occur
within the precursor of a strong, accelerating shock. The diffusive streaming
of accelerated CRs is expected to be so strong in this region that a purely
growing MHD mode appears with a growth rate that is, at least in the outer part
of the precursor, larger than the growth rate of the well-known resonant
Alfv\'enic mode. In addition it was demonstrated \citep{Rbell12} that CRs form
filamentary structures in the
precursor due to their self-generated magnetic field. This filamentation
results in the growth of a long-wavelength instability.  It is expected that
due to these instabilities the external magnetic field $B_\mathrm{ISM}$ is
amplified within the entire precursor structure up to $B \gg B_\mathrm{ISM}$,
since the main effect is produced by the most energetic CRs with momenta $p
\sim p_\mathrm{max}$ which populate the whole precursor diffusively during their
acceleration.

The Alfv\'en wave excitation in the shock precursor \citep{bell78} corresponds
to an additional unstable mode. It leads, in particular, to a high level of
resonantly scattering waves. Therefore, the overall field amplification will be
the result of all instabilities operating in the precursor.

These mechanisms should lead to the Bohm limit
$\kappa(p)=\kappa_\mathrm{Bohm}(p)$ for CR diffusion which is achieved when the
condition $E_\mathrm{w}(k)=E_B$ is is reached.

CRs with the momentum spectrum $N\propto p^{-\gamma}$ produce an Alfv\'en wave
spectrum $E_\mathrm{w}\propto k^{\gamma -2}\propto p ^{2-\gamma}$, where $k$
and $p$ are approximately connected by the relation
$k=1/\rho_\mathrm{B}$.  It can be represented in the form
\begin{equation}
E_\mathrm{w}(p)=E_\mathrm{m} \times (p/p_\mathrm{max})^{2-\gamma},
 \label{eq4}
\end{equation}
where $E_\mathrm{m}=E_\mathrm{w}(p_\mathrm{max})$.

The amplified magnetic field $B$ in the case of a steep CR spectrum is
produced by CRs with energy $p \sim m_\mathrm{p}c$ because it is
these CRs that provide the main contribution to the overall CR energy content;
here $m_\mathrm{p}$ denotes the proton rest mass. Therefore the
Bohm limit condition $E_\mathrm{w}(p)=E_\mathrm{B}$ can be fulfilled for $p
\sim m_\mathrm{p}c$, whereas for CRs with $p
  \approx 10^6 m_\mathrm{p}c$ one would have $E_\mathrm{w} \approx
10^{(2-\gamma)\times 6} E_\mathrm{B} \approx 0.1 E_\mathrm{B}$, for $\gamma
=2.2$. This means
that the CR diffusion coefficient of such high-energy particles would
exceed the Bohm limit by roughly one order of magnitude. As a consequence, this
amplified field would play almost no role for CRs with high momenta $p
\gg m_\mathrm{p}c$, and therefore the maximal CR momentum that can be
achieved would be considerably lower than $10^{6} m_\mathrm{p}c$.  This
is in contradiction to the assumption of MC12 that
$\kappa(p)=\kappa_\mathrm{Bohm}(p)$ for the entire range of CR momenta $p\le
10^{6} m_\mathrm{p}c$ considered.

This argument demonstrates that a considerable increase of the maximum CR
momentum due to magnetic field amplification is expected only in the case of a
hard CR spectrum $N\propto p^{-\gamma}$ with $\gamma\le 2$, where the CRs with
the highest momenta $p \sim p_\mathrm{max}$ provide the main contribution to the
overall CR energy content. Fortunately, such a spectrum is expected to be
produced by SN shocks for injection rates that considerably exceed those used
by MC12.

\section{CR and \gr production in a clumpy ISM}

The physics aspect which is not included in the present kinetic and
(``renormalized'') spherically symmetric model is an essential inhomogeneity of
the ambient ISM on spatial scales that are smaller than the SNR radius. This
inhomogeneity is not the result of the progenitor star's evolution towards the
final supernova explosion, for example in the form of a wind and a
corresponding modification of the circumstellar environment. It is rather an
inherent nonuniformity of the average ISM on account of (i) the interplay
between its radiative heating by the diffuse galactic UV field and the
radiative cooling of the gas \citep{field1965} and (ii) the stochastic
agitation of the ISM by the mechanical energy input and gas heating from
supernova explosions \citep{cox74,mckee77}. The first effect is a thermal
instability and thus a mechanism for small-scale cloud formation in the ISM
driven by runaway radiative cooling \citep{field1969}. Specifically the
balance between line-emission cooling and
gas heating due to the ultraviolet background radiation leads to two thermally
stable equilibrium ISM phases \citep{wolfire2003}. One of them is the so-called
warm interstellar medium with a typical gas number density $N_\mathrm{g1} \sim
0.1$~cm$^{-3}$ and temperature $T_1\approx 8000$~K, the other one a cold
neutral medium with $N_\mathrm{g2}\sim 10$~cm$^{-3}$ and $T_2\approx 100$~K.
According to simulations the scale of dense clouds is typically
$l_\mathrm{c}=0.1$~pc \citep[e.g.][]{audit08,audit10,inoue12}
  \footnote{Gravitationally bound molecular clouds are not considered here,
    because they are large-scale massive elements of the ISM. If present, they
    would require an individual treatment.}. The second effect is a general
compressible turbulence of the ISM, at least on scales in excess of $\sim
1$~pc, driven by the Galactic supernova explosions. According to
MHD-simulations it involves high- and low-temperature gas components out of
ionization equilibrium on all larger scales \citep{avillez05}, for a
  review see \citet{breitschwerdt12}. While the overall picture of the ISM is
  clearly not simple, the evolution of a young SNR like that of Tycho's SN will
  encounter a single realization of the stochastic ensemble of density
  fluctuations. For the energy spectrum of energetic particles, accelerated at
  the blast wave, small-scale high-density fluctuations of the ISM play the
  most conspicuous role because they produce mainly particles with energies far
  below the cutoff energies for a uniform circumstellar medium. To estimate the
  spectral changes due to upstream density variations in an analytical model,
  the typical ISM is therefore treated here as a generalized
two-phase medium, composed of a pervasive warm/hot ISM (phase I) -- called here
for brevity ``warm'' ISM -- and small-scale dense clouds (phase II), embeded in
this warm ISM.

In order to determine the specific properties of the CRs and their nonthermal
emission in the case of such a generalized two-phase ISM the latter is
approximated here in a simple form, as a uniform warm phase with gas number
density $N_\mathrm{g1}$ plus an ensemble of small-scale dense clouds with gas
number density $N_\mathrm{g2}$. The warm diluted ISM phase is assumed to have a
volume filling factor $F_1\approx 1$, whereas the clouds occupy a small
fraction of space with filling factor $F_2\ll 1$. It is in addition assumed
that most of the gas mass is contained in the warm phase, which means that
$F_1N_\mathrm{g1}\gg F_2N_\mathrm{g2}$. Then
the SN shock propagates in the two-phase ISM without essential changes compared
with the case of a purely homogeneous ISM with number density $N_\mathrm{g1}$
\citep{inoue12}. Therefore it produces inside the phase~I of the ISM roughly
the same amount of CRs and nonthermal emission as in the case of a homogeneous
ISM. Then one has to estimate the additional contribution of the clouds in
order to determine the overall spectrum of CRs.

The large-scale SN blast wave, interacting with each single cloud, produces a
pair of secondary transmitted and reflected shocks. The reflected shock
propagates in the warm isobaric ISM already heated by the blast
wave. Due to this fact its Mach number is quite low and therefore its
contribution to the overall CR production can be neglected.

The size $R_\mathrm{s2}=l_\mathrm{c}/2$ and the speed $V_\mathrm{s2} \approx
(N_\mathrm{g1}/N_\mathrm{g2})^{1/2}V_\mathrm{s} \sim 10^{-1}
V_\mathrm{s}$ of the transmitted shock are both considerably smaller
than the corresponding values of the SN blast wave. Therefore, according to
Eq.(\ref{eq1}), and using $A=10$ for both shocks, the maximum momentum of CRs
produced inside the cloud
\begin{equation}
  p_\mathrm{max2}=R_\mathrm{s2}V_\mathrm{s2}/(R_\mathrm{s}V_\mathrm{s}) p_\mathrm{max1}
 \label{eq5}
\end{equation}
is much smaller than the maximal momentum $p_\mathrm{max1}$ of the CRs
produced by the SN shock in the warm ISM: $p_\mathrm{max2}\ll
p_\mathrm{max1}$.  Since the ram pressure $\rho V_\mathrm{s}^2$ is expected to
be the same in both cases, and since the amplified magnetic field pressure
reaches roughly the same fraction of the ram pressure, the magnetic field values
are roughly the same in these two cases: $B_2 \sim B_1$.

An estimate of the CR spectrum produced by the transmitted shock starts from
the expression for the CR distribution function
\begin{equation}
f=\frac{q\eta N_\mathrm{g}}{4\pi p_\mathrm{inj}^3}
\left(\frac{p}{p_\mathrm{inj}}\right)^{-q}, 
\label{eq6}
\end{equation}
which is valid at all momenta $p \geq p_\mathrm{inj}$ up to the cutoff momentum
$p_\mathrm{max}$ in the case of an unmodified shock, and roughly valid within the
momentum range $p_\mathrm{inj}< p < 10 m_\mathrm{p}c$ of the subshock in
the case of a modified shock. Using this expression for the case of the SN blast
wave and for the transmitted shock, the ratio of the two corresponding distribution
functions can be found as
\begin{equation}
\frac{f_2}{f_1}=\frac{q_2}{q_1 }
\left(
\frac{ N_\mathrm{g2}}{N_\mathrm{g1}}
\right)^{1-(q_2-3)/2}
\left(
\frac{p}{p_\mathrm{inj1}}
\right)^{q_1-q_2}, 
\label{eq7}
\end{equation}
in $p_\mathrm{inj} < p \leq 10 m_\mathrm{p}c$,
where also the expression for the injection momentum \citep{byk96}
%
%
%
\begin{equation}
p_\mathrm{inj} \approx m_\mathrm{p} V_\mathrm{s}.
\label{eq8}
\end{equation}

was used.

According to the calculation from section 2, the power
law spectrum with the index $q_1\approx 4.3$, determined by the subshock
compression ratio, extends up to the CR momenta $p\sim 10m_\mathrm{p}c$.

Since the cutoff momentum $p_\mathrm{max2}\ll p_\mathrm{max1}$ of the CR
spectrum, produced by the transmitted shock, is much lower than the
corresponding value for the blast wave, $p_\mathrm{max1}\approx 10^6
m_\mathrm{p}c$, one can neglect the modification of the transmitted shock. This
leads to $q_2\approx 4$.

Taking into account that the \gr production is proportional to the gas density,
for $\epsilon_{\gamma} \gtrsim 1$~GeV one can write the relation between
the fluxes of \grs produced due to the two shocks considered:
\begin{equation}
F_{\gamma2}(\epsilon_{\gamma})=aF_2F_{\gamma1}(\epsilon_{\gamma})
\exp(-\epsilon_{\gamma}/\epsilon_{\gamma \mathrm{max2}}).
\label{eq9}
\end{equation}
Here the factor $a$ is determined by the expression
\begin{equation}
a=\frac{q_2}{q_1}
\left(\frac{N_\mathrm{g2}}{ N_\mathrm{g1}}\right)^{1.5}
\left(
\frac{10c}{V_\mathrm{s}}
\right)^{0.3}
\label{eq10}
\end{equation}
and the \gr cutoff energy is $\epsilon_{\gamma \mathrm{max2}} \sim
0.1p_\mathrm{max2}c$ since on average the energy of the \grs resulting from
inelastic proton-proton collisions is about one tenth of the proton energy:
$\epsilon_{\gamma} \sim 0.1 pc$.

Substituting the values of the SN shock speed $V_\mathrm{s}=5000$~km/s 
\citep{katsuda10,vbk08}, the number density for the warm phase of the ISM
$N_\mathrm{g1}=0.25$~cm$^{-3}$, as well as suitable fit parameter values for
the cold phase of the ISM in the form of $N_\mathrm{g2}\approx
23N_\mathrm{g1}=6$~cm$^{-3}$ and $F_2=0.005$, results in
$p_\mathrm{max2}=10^{-3} p_\mathrm{max1}=10^3m_\mathrm{p}c$ and $a\approx
800$. It is noted here that such parameter values for the cold ISM phase
correspond rather well to the results of numerical modelling of the two-phase
ISM \citep{inoue12}.

Then the flux of \grs produced inside the clouds can be written in the form
\begin{equation}
F_{\gamma2}(\epsilon_{\gamma})=4F_{\gamma1}(\epsilon_{\gamma})
\exp(-\epsilon_{\gamma}/100\mathrm{GeV}).
\label{eq11}
\end{equation}

The total \gr flux $F_{\gamma}=F_{\gamma1}+F_{\gamma2}$, expected from Tycho's
SNR for a two-phase ISM is shown in Fig.\ref{f1}. One can see that it fits the
existing data in a satisfactory way. Note that the considerable increase (by a
factor of 5) of the \gr emission at energies $\epsilon_{\gamma}< 100$~GeV over
and above the case of a purely homogeneous ISM is due to the contribution of
clumps which contain only 10\% of the ISM mass.

\section{Discussion}

Small-scale dense clumps of sizes $l_\mathrm{c}\ll 0.1R_\mathrm{s}$ can be also
produced by the accelerating CRs themselves within the precursor as the result
of the so-called acoustic instability \citep{dorfi85,drury84,ber86}.

Small-scale bright structures of angular size $10''$ were recently detected in
nonthermal X-rays \citep{eriksen11}. For a source distance $d=3.8$~kpc the
corresponding spatial size is $l\approx 0.2$~pc, which would be roughly
consistent with the sizes of the expected clumps. However, the acceleration in
such clouds is not expected to reach electron energies in the TeV range which
could lead to synchrotron X-ray emission. Therefore these small-scale X-ray
structures can not be considered as an indication that dense gas clumps of size
$l_\mathrm{c}\sim 0.1$~pc indeed exist inside Tycho's SNR. Their existence
rather derives from the general properties of the ISM, as discussed above.

The contribution of dense gas clumps in different kinds of emission can be
roughly estimated as follows. First, consider the thermal X-ray emission.
Since, besides other factors, the flux of thermal X-ray emission
$F_\mathrm{X}\propto M_\mathrm{g}N_\mathrm{g}$ is proportional to the gas
density $N_\mathrm{g}$ and the total gas mass $M_\mathrm{g}$ of the source, we
have $F_\mathrm{X2}/F_\mathrm{X1}=
(F_2/F_1)(N_\mathrm{g2}/N_\mathrm{g1})^2\approx 2.6$.  Since the temperature
difference in the two gas phases is not an essential factor for the soft X-rays
with energies below 2~keV \citep[e.g.][]{hamilton83} we conclude that the soft
thermal X-ray emission should be dominated by the contribution of dense gas
clumps. In the hard X-ray range above 2~keV, on the other hand, the luminosity
is sensitive to the gas temperature, roughly as $F_\mathrm{X}\propto
T^{2.1}\propto V_\mathrm{s}^{4.2}$ \citep[e.g.][]{hamilton83} which make the
contribution of dense gas clumps relatively small due to their lower
temperature. The expected luminosity of individual clumps is considerably
higher (by a factor of about 500) compared with the surrounding diluted gas of
the same volume. However each instrument sees the remnant in projection.
Therefore the expected ratio of X-ray fluxes from the projection volume
$V=\pi\rho^2 L$ containing the clump to the nearby one of the same size which
does not contain the clump is $r=(V+500V_\mathrm{c})/V$, where
$V_\mathrm{c}=\pi l_\mathrm{c}^3/6$ is the clump volume, $\rho$ is the
cross-section of the volume, and $L$ is the line of sight length. A maximal
value of this ratio $r\approx 1+330l_\mathrm{c}/L$ is achieved for
$\rho=l_\mathrm{c}/2$.  It follows from this expression that the contrast of
X-ray emissions varies from $r\approx 4$ for the central part of the remnant
where $L\approx 10$~pc to about $r\approx 25$ at the limb region with $L\approx
1.4$~pc. Therefore we conclude that these clumps could be detected in soft
X-rays from the limb regions. In their study with Chandra \citet{cassam07}
  observe some small contribution of thermal X-rays from the regions occupied
  by the swept-up ambient gas. They concluded that it is not clear whether the
  faint lines or other residual emission comes from the ejecta or whether they
  arise from shocked ambient medium. In line with the idea of the present paper
  it is suggested here that this emission comes from small-scale clouds in the
  surrounding ISM. The difficulty is of course that the
X-ray emission of Tycho's SNR is dominated by the nonthermal X-ray
emission. In the context of X-ray emission also the observed
  irregularities in the blast wave position around the remnant
  \citep[e.g.][]{warren05} could at least be partly due to the interaction with
  ambient clouds.

Secondly, the contribution of dense gas clumps to the synchrotron emission of
SNRs is estimated here. From their measurements of the variations in the
  expansion parameter in the radio range and comparison with X-ray features
  \citet{reynoso97} suggested the presence of ambient clouds, shocked by the
  blast wave. Since the radio synchrotron emission is produced by electrons
with energies less then 1~GeV, Eq.(\ref{eq7}) with $p=m_\mathrm{p}c$ is used in
order to estimate the ratio of the total synchrotron fluxes originating within
the two gas phases: $F_{\nu2}/F_{\nu1}\propto (F_2/F_1)(f_2/f_1)\approx 0.1$.
This shows that the contribution of dense gas clumps to the radio emission is
expected to be small.  The luminosity of individual clumps compared with the
neighbouring region of the same volume is about 15. Therefore one should be
able to detect the clumps with an instrument of corresponding angular
resolution from the limb region where the contrast is expected to be $r\approx
2.9$. The synchrotron X-ray energy flux scales as $F_\mathrm{X}\propto
FK_\mathrm{ep}N_\mathrm{g}V_\mathrm{s}^2\propto F$. Therefore the expected
X-ray flux from all the clumps within the SNR, $F_\mathrm{X2}\approx
(F_2/F_1)F_\mathrm{X1}\approx 5\times10^{-3}F_\mathrm{X1}$, is small compared
with the total flux $F_\mathrm{X1}$.

The dense clumps of size $l_\mathrm{c} \approx 0.1$~pc have an angular size
of about $6''$. Structures of such sizes (or
even smaller) near the outer shock of Tycho's SNR have been studied in optical
H$\alpha$ emission \citep[e.g.][]{ghavamian2000, lee2010}. It is difficult to
conclude whether the dense clumps can be detected in optical emission,
  even though the lack of smoothness of the optical filaments points in
  this direction.

It should also be noted that there are some observations which show that
Tycho's SNR is expanding into an ISM with large scale density gradients and is
possibly interacting with a dense ambient medium towards the northern direction
\citep{lee2004}. If confirmed this makes the overall picture of this SNR even
more complicated.  It is nevertheless suggested here that the spherically
symmetric approach assuming a uniform ISM on a large scale is roughly valid.

  According to the above considerations, the excess of \gr GeV emission above
  the level expected in the case of a purely homogeneous ISM due to the
  contribution of  small-scale interstellar clouds is inevitably expected
  in the case of type Ia SNe situated in a relatively dense ISM inside the
  Galactic disk.  It is not clear whether such an effect is expected in the
  case of a SNR situated in a much more rarefied ISM, like SN~1006, in the
  uppermost part of the Galactic gas disk. On a speculative basis, similar
  effects may also take place in Cassiopeia~A, where the observed \gr spectrum
  \citep{CasFermi10} looks very similar to that of Tycho's SNR. The difference
  would be that in the case of Cassiopeia~A numerous observed knots from the
  supernova ejecta might take over the role of pre-existing interstellar
  clumps.

\section{Summary}

The \gr spectrum of Tycho's SNR, consistent with the measurements by Fermi and
VERITAS, is proposed to be the superposition of two spectra: the first
part, extended up to about $100$~TeV, is produced by the SN blast wave within
the dilute ``warm'' phase of the ambient ISM, whereas the second part, with a
cutoff at about $100$~GeV, originates in dense clouds embedded in this warm ISM.
The remarkable connection between CR production and the physical nature of the
Galactic ISM becomes evident through the characteristics of the spatially
integrated \gr emission of the SNR sources.

\acknowledgments
This work has been supported in part by the Department of Federal Target
Programs and Projects (Grant 8404), by the Russian Foundation for Basic
Research (grants 10-02-00154 and 11-02-12193) and by the Council of the
President of the Russian Federation for Support of Young Scientists and Leading
Scientific Schools (project NSh-1741.2012.2). EGB acknowledges the
hospitality of the Max-Planck-Institut f\"ur Kernphysik, where part of this
work was carried out.

\end{document}